\documentclass[twocolumn,amsmath,amssymb,prd]{revtex4}
\usepackage{graphicx}

\def\al{\alpha}
\def\be{\beta}
\def\ga{\gamma}
\def\de{\delta}

\def\th{\theta}

\def\la{\lambda}

\def\si{\sigma}

\def\ta{\tau}

\def\De{\Delta}

\def\ab{{\al\be}}

\def\etal {{\it et al.}}
\newcommand{\beq}{\begin{equation}}
\newcommand{\eeq}{\end{equation}}
\newcommand{\bea}{\begin{eqnarray}}
\newcommand{\eea}{\end{eqnarray}}
\newcommand{\bse}{\begin{subequations}}
\newcommand{\ese}{\end{subequations}}

\def\ni{\noindent}
\def\sF#1#2{{\textstyle{{#1}\over{#2}}\,}}
\newcommand{\BB}{\big}
\newcommand{\nn}{\nonumber}

\def\ol#1{\overline{#1}}

\usepackage{amsmath}
\usepackage{amsfonts}
\usepackage{amssymb}

\newcommand{\bM}{\begin{pmatrix}}
\newcommand{\eM}{\end{pmatrix}}

\newcommand{\ha}{\sF{1}{2}}

\newcommand{\Tr}{\text{Tr}}

\newcommand{\B}{\Big}
\newcommand{\R}{\text{Re}}
\newcommand{\I}{\text{Im}}

\def\e{\varepsilon^m}
\def\ed{\varepsilon^{m*}}
\def\a{(a_L)^{0}} 
\def\ad{(a_L)^{0*}} 
\def\M#1#2{({\cal M}^{(1)}_{#1})_{#2}}

\begin{document}

\title{Correspondence between nonstandard interactions and CPT violation \\ in neutrino oscillations}
\author{Jorge S. D\'iaz}
\affiliation{Institute for Theoretical Physics, Karlsruhe Institute of Technology, 76128 Karlsruhe, Germany}

\begin{abstract}
The experimental signatures of nonstandard neutrino interactions are shown to be equivalent to CPT violation in neutrino oscillations.
This result leads to a correspondence in the study of these two descriptions that can be used to constrain the relevant parameters of one formalism by using the available bounds on parameters of the other.
The correspondence is illustrated and explicitly used to determine first bounds on previously unexplored parameters in both formalisms.
\end{abstract}

\maketitle

\section{Introduction}

Neutrino oscillations have been experimentally confirmed using accelerator, atmospheric, reactor, and solar neutrinos \cite{PDG2014}.
These observations indicate the remarkable interpretation of massive neutrinos.
The minimal model extending the Standard Model (SM) to accommodate neutrino masses has been verified with great precision in all the experiments above \cite{PDG2014}.
Even though to date the absolute mass scale of neutrinos remains unknown,
the interferometric nature of neutrino oscillations has allowed us to measure the effects of their minute masses in the form mass-squared differences.
This unexpected behavior within the context of the SM has motivated the search for unconventional effects that could manifest their minuscule effects in neutrino oscillations.

These so-called exotic scenarios beyond the SM have become very active fields and their study gave rise to an interesting incidental program of experiments originally designed to measure the parameters of the three-neutrino massive model.
Some of these ideas include the search for nonstandard neutrino interactions (NSI) \cite{MSW,NSI0},
Lorentz and CPT violation in neutrinos \cite{KM2004},
long-range interactions \cite{LRInu},
large extra dimensions \cite{ExtraDnu},
and 
sterile neutrinos \cite{SterileNus}.
Regarding the first two exotic scenarios mentioned above,
the physical motivations are completely independent;
nonetheless,
the observable effects that they would produce in experiments can be related.
This means that although they remain as independent modifications of the conventional physics,
the experimental constraints of one can help to constrain the other.

In this work, a correspondence between matter NSI and CPT violation in neutrinos is presented,
which allows relating the parameters in these two formalisms.
This paper is organized as follows.
Matter NSI and their effects in neutrino oscillations are described in Sec. \ref{Sec.NSI},
while the corresponding effects of CPT violation are presented in Sec. \ref{Sec.CPT-}.
In Sec. \ref{Sec.NSI_CPT} the correspondence between matter NSI and CPT violation is discussed 
and the explicit application of a perturbative method is illustrated.
The proposed correspondence is used in sections \ref{Sec.bounds1} and \ref{Sec.bounds2} to relate and determine new bounds on the parameters of both formalisms.
Sec. \ref{Sec.future} describes a brief discussion on the sensitivity prospects for future experiments.

\section{Nonstandard interactions}
\label{Sec.NSI}

Neutrino interactions at low energies can be effectively described by four-fermion vertices containing two neutrino states.
In particular,
nonstandard interactions of neutrinos with quarks and electrons in a material medium can be described by the effective Lagrangian \cite{MSW,NSI0}
\beq\label{L_NSI}
\mathcal{L}_{\text{NSI}} 
=
-2\sqrt{2} \,G_{F} 
\BB(\ol{\nu}_\al \ga^\mu \nu_\be \BB)
\BB(\varepsilon_{\ab}^{ff'\!P}\,\ol{f}_P \ga_\mu f_P' \BB)
+\text{h.c.}
\eeq
where $G_F$ is the Fermi constant, $P=L,R$, and the strength between neutrino states $\nu$ of flavors $\al$ and $\be$ and the $L$-handed ($R$-handed) components of the fermions $f$ and $f'$ is parametrized by the factors $\varepsilon_{\ab}^{ff'\!L}$ ($\varepsilon_{\ab}^{ff'\!R}$).
For $f\neq f'$, the NSI produce charged-current type effects that modify the neutrino production and detection.
These effects are controlled by NSI parameters denoted $\varepsilon_{\ab}^{\text{s}}$ and $\varepsilon_{\ab}^{\text{d}}$ referring to the source and detector, respectively.
Remarkably, these parameters can trigger neutrino flavor change even in the limit of no propagation through the so-called {\it zero-distance effect} \cite{zeroLeffect} and several proposals for the experimental study of these parameters have been considered \cite{NSIreview}.
In the present work, we are interested on NSI parameters that modify the neutrino propagation through matter, 
which occurs in the case $f=f'$ in the form of NSI neutral-current type effects.

In vacuum, 
the effective hamiltonian describing neutrino oscillations can be written in terms of the mass-squared differences $\De m^2_{21}$, $\De m^2_{31}$, and the neutrino energy $E$ in the form
\beq\label{H0}
H_0 = \frac{1}{2E}\,U
\bM 
0 & 0 & 0 \\
0 & \De m^2_{21} & 0 \\
0 & 0 & \De m^2_{31} \\ \eM 
\,U^\dag,
\eeq
where $U$ is the PMNS matrix that parametrizes the mixing between flavor and mass eigenstates in terms of four constant parameters: 
three mixing angles and one CP phase \cite{PMNS}; 
Majorana phases could also be included,
although they are unobservable in neutrino oscillations.
For neutrino experiments on Earth, 
in particular long-baseline beam experiments, 
the propagation through matter can play a crucial role \cite{MSW}.
Forward scattering of neutrinos with the particles that they encounter as they propagate in a dense medium can be modelled as a constant contribution to the hamiltonian in the form $H=H_0+H_\text{M}$, where
\beq\label{H_M}
H_\text{M} = \sqrt2\, G_F n_e
\bM 
1 & 0 & 0 \\
0 & 0 & 0 \\
0 & 0 & 0 \eM .
\eeq
This matter potential arises from charged-current interactions with electrons in the medium of number density $n_e$.
Neutral-currents modify the three neutrino flavors by the same amount;
being proportional to the identity in flavor space,
these interactions produce no effects in the oscillations of three neutrinos so they can be disregarded.

In a similar fashion, 
the vector component of the NSI \eqref{L_NSI} for $f'=f$ modifies neutrino propagation through matter,
in which case the combination 
\beq
\e_{\ab}=\sum_{f=u,d,e}(\varepsilon_{\ab}^{ffL}+\varepsilon_{\ab}^{ffR})\frac{n_f}{n_e}
\eeq
controls the observable effects.
In the last expression, the number density of the relevant fermion of type $f$ is denoted by $n_f$ 
and we have considered that neutrinos propagate through an unpolarized medium.
The physical effect of the parameters $\e_{\ab}$ is a modification of the neutrino dispersion relation similar to the matter effect \eqref{H_M} affecting all flavor components independently.
The corresponding effective hamiltonian takes the general form
\beq\label{H_NSI}
H_{\text{NSI}} = \sqrt2\, G_F n_e
\bM 
\e_{ee} & \e_{e\mu} & \e_{e\ta} \\
\ed_{e\mu} & \e_{\mu\mu} & \e_{\mu\ta} \\
\ed_{e\ta} & \ed_{\mu\ta} & \e_{\ta\ta} \eM ,
\eeq
where the flavor structure guarantees the hermiticity of the hamiltonian.
For the propagation of antineutrinos,
the corresponding NSI hamiltonian is obtained by the replacement 
$\sqrt2\, G_F n_e\,\e_{\ab}\to-\sqrt2\, G_F n_e\,\ed_{\ab}$. 
Notice that due to the absence of antimatter in the medium, 
spurious CP- and CPT-violating effects can appear when neutrinos propagate in matter \cite{IntExtCPTv}.

The full hamiltonian describing neutrino oscillations through a dense medium in the presence of NSI can be written as
\beq\label{H}
H = H_0+H_{\text{M}}+H_{\text{NSI}}.
\eeq
The study of the parameters $\e_\ab$ in the last term is addressed by the diagonalization of the full hamiltonian \eqref{H}.
Given the mixed energy dependence of the different terms in this hamiltonian,
the mixing angles will be energy-dependent.
This phenomenon is well known from the study of solar neutrinos,
in which case the matter potential is given in terms of the electron density in the solar core.
Moreover, 
the eigenvalues of the hamiltonian will exhibit an unconventional dependence of the neutrino energy that will differ from the vacuum behavior that makes the oscillation phase proportional to $E^{-1}$.
Different approaches have been implemented for the search of matter NSI using atmospheric \cite{NSI_atm}, beam \cite{NSI_beam}, reactor \cite{NSI_reactor}, and solar \cite{NSI_solar} experiments.
Distinct methodologies have been explored for the direct study of the NSI parameters in neutrino oscillations, 
including approximation techniques that can be implemented in some experimental configurations.
In particular, 
given the remarkable success of the model of three massive neutrinos,
NSI are usually considered as sub-leading effects that could modify the conventional description of neutrino oscillations \cite{NSIreview,NSI_atm,NSI_beam,NSI_reactor,NSI_solar}.
To date,
most of the constraints on the NSI parameters $\e_{\ab}$ have been obtained from phenomenological studies \cite{Biggio:2009}.
Two direct experimental studies have been performed by the MINOS \cite{NSI_MINOS} and Super-Kamiokande \cite{NSI_SuperK} collaborations.
Hereafter,
we only consider bounds obtained from experimental studies.

\section{CPT violation}
\label{Sec.CPT-}

Neutrino oscillations are natural interferometers that offer great sensitivity to search for new physics, 
including deviations from exact Lorentz invariance.
The potential breakdown of one of the most fundamental symmetries of modern physics has been mostly motivated by string-theory scenarios \cite{SBS_LV}. 
Systematic searches of deviations from Lorentz symmetry have been implemented using the so-called Standard-Model Extension (SME) \cite{SME1,SME2}.
These modern tests of Lorentz invariance span a wide range of fields, 
whose experimental results are tabulated in Ref. \cite{tables}.
In the neutrino sector \cite{KM2004}, 
different techniques have been developed for weak decays \cite{DKL2013,JSD_SBD,JSD_DBD} and astrophysical neutrinos \cite{Astro_nus}; 
nevertheless, 
neutrino oscillations using short- and long-baseline experiments have shown to be very sensitive probes of unique forms of Lorentz violation that lead to neutrino mixing \cite{KM2004_SB,DKM2009}.

There exists a subset of Lorentz-violating operators that also break CPT invariance in the fermion sector.
Restricting attention only to renormalizable Dirac couplings in the theory,
these CPT-odd operators are written in the form \cite{KM2004,KM2012}
\beq\label{L_CPT-}
\mathcal{L}_{\text{CPT--}} = 
-\ha a_{\ab}^\la \ol\psi_\al\ga_\la\psi_\be
-\ha b_{\ab}^\la \ol\psi_\al\ga_5\ga_\la\psi_\be
+ \text{h.c.},
\eeq
with the flavor indices taking the values $\al,\be=e,\mu,\ta$ 
and where the coefficients $a_{\ab}^\la$ and $b_{\ab}^\la$ control the impact of vector and pseudo-vector couplings, 
respectively.
It is important to emphasize that coordinate invariance is preserved and only particle Lorentz invariance is broken \cite{JSDreview2014}.
We remark in passing that Majorana couplings can also be included in the lagrangian \eqref{L_CPT-},
which can mix neutrinos and antineutrinos \cite{KM2004,DKM2009}.
These Majorana couplings produce direction-dependent effects and have been experimentally constrained using beam \cite{RebelMufson}, reactor \cite{LV_DC2}, and double-beta-decay experiments \cite{JSD_DBD}.
From the lagrangian \eqref{L_CPT-}, 
the observable effects on left-handed neutrinos are controlled by the combinations
\beq\label{aL}
(a_L)_{\ab}^{\la} = (a + b)_{\ab}^{\la},
\eeq
which are constant hermitian matrices in flavor space that modify the conventional vacuum hamiltonian through an observer-independent structure of the form 
$(a_L)_{\ab}^{\la}\,\hat p_\la$ \cite{KM2004,KM2012}.
The corresponding hamiltonian for antineutrinos is obtained after the coefficients \eqref{aL} are substituted by the right-handed coefficients 
$(a_R)_{\ab}^{\la}=(a - b)_{\ab}^{\la}=-(a_L)_{\ab}^{\la*}$.
The spacetime index $\la=0,1,2,3$ explicitly exhibits the potential breakdown of rotational invariance,
where $\hat p^\la=(1;\hat{\pmb{p}})$ depends on the neutrino direction of propagation $\hat{\pmb{p}}$.
Nonetheless, 
in the present work we are only interested in the isotropic component $(a_L)_{\ab}^{0}$.

The explicit form of the CPT-violating modification of the neutrino hamiltonian can be written as
\beq\label{H_CPT}
H_\text{CPT--} = 
\bM 
\a_{ee} & \a_{e\mu} & \a_{e\ta} \\
\ad_{e\mu} & \a_{\mu\mu} & \a_{\mu\ta} \\
\ad_{e\ta} & \ad_{\mu\ta} & \a_{\ta\ta} \eM ,
\eeq
where the components of $\a_{\ab}$ completely characterize independent deviations from CPT invariance.
The full hamiltonian describing oscillations of CPT-violating neutrinos can be written as
\beq\label{H2}
H = H_0+H_{\text{M}}+H_{\text{CPT--}},
\eeq
where we have included the matter hamiltonian for completeness.
The diagonalization of this hamiltonian can then be used to study the coefficients $\a_{\ab}$ in oscillation experiments.
For this purpose,
perturbative methods have been developed for short- and long-baseline experiments \cite{KM2004_SB,DKM2009}.
Many experimental studies have been performed to constraint these coefficients by 
Double Chooz \cite{LV_DC},
IceCube \cite{LV_IceCube},
LSND \cite{LV_LSND},
MiniBooNE \cite{LV_MiniBooNE1,LV_MiniBooNE2},
MINOS \cite{LV_MINOS_ND1,LV_MINOS_FD,LV_MINOS_ND2}, and
Super-Kamiokande \cite{LV_SuperK}.
Similarly,
sensitivity studies have been performed for the future JUNO experiment \cite{LV_JUNO}.
The hamiltonian \eqref{H2} has also been used to implement alternative models for neutrino oscillations \cite{LVmodels}.

\section{NSI-CPT-violation correspondence}
\label{Sec.NSI_CPT}

Comparing the effects at the hamiltonian level of NSI \eqref{H_NSI} and CPT violation \eqref{H_CPT} we can directly make the parameter correspondence 
\beq\label{CPT=NSI}
\a_{\ab} \leftrightarrow \sqrt2\, G_F n_e\, \e_{\ab}.
\eeq
This expression establishes an equivalence between the observable effects introduced by the matter NSI parameters $\e_{\ab}$ and the coefficients for CPT violation $\a_{\ab}$.
The modified neutrino oscillation probabilities in both scenarios will take the same form when describing CPT-violating neutrinos modelled by the theory \eqref{L_CPT-} or when introducing matter NSI described by the theory \eqref{L_NSI};
the relation \eqref{CPT=NSI} shows the connection between the two formalisms.
This result implies that the experimental bounds obtained in one formalism could be translated into bounds for the other.

It is important to emphasize that even though the relation \eqref{CPT=NSI} allows us to relate the parameters of matter NSI and CPT violation in neutrino oscillations, 
the underlying physics controlled by the corresponding parameters remains inequivalent.
In fact,
matter NSI require that neutrino propagate through matter to be observable,
whereas the effects of CPT violation are independent of the medium.
In both cases,
there exists an effective refractive index that modifies the neutrino dispersion relation.
For matter NSI, 
this index of refraction arises due to the unconventional interactions with electrons and quarks in the medium, 
which act like a condensate that alters the neutrino propagation.
On the other hand,
the index of refraction in the CPT-violating scenario corresponds to the existence of an intrinsic background field that isotropically permeates the vacuum.
The nature of this an other background fields has been extensively studied for theories with Lorentz invariance violation.
For an updated review, 
see Ref. \cite{TassonReview} and references therein. 

Now that we have established a direct relationship between the matter NSI parameters and the coefficients that control CPT violation,
we can borrow the methods from one framework to apply it to the other.
Caution is necessary for a correct use of the relation \eqref{CPT=NSI} because there might exist situations in which the relationship between parameters is invalid due to the nature of the analysis to extract limits on the parameters.
For instance, 
some bounds on NSI have been obtained assuming a particular matter composition so the corresponding electron density must be treated with care.
Conversely,
some bounds on coefficients for CPT violation have been obtained using neutrinos that propagate in different media; 
hence, again the correct interpretation of the matter density must be taken into account.

Many of the constraints on CPT-violating neutrinos have used perturbative methods,
treating CPT violation as a sub-leading effect over the conventional mass-driven oscillations,
an approach also implemented in the study of matter NSI.
For instance, for long-baseline experiments the far detector is located at a distance from the source that maximizes the conventional oscillation effects.
A power-series expansion of the oscillation probability in the form 
$P_{\nu_\be\to\nu_\al}=P_{\nu_\be\to\nu_\al}^{(0)}+P_{\nu_\be\to\nu_\al}^{(1)}+P_{\nu_\be\to\nu_\al}^{(2)}+\ldots$ 
allows identifying the different terms in the series as a function of the sub-leading physical process added to the conventional description.
As expected,
the first term in the series is the conventional neutrino oscillation probability driven by neutrino mass-squared differences.
Using the results in Ref. \cite{DKM2009},
this conventional probability can be written as $P_{\nu_\be\to\nu_\al}^{(0)}=|S^{(0)}_{\ab}|^2$,
where the zeroth-order oscillation amplitude for neutrinos that propagate a distance $L$ is given by
\beq
S^{(0)}_{\ab} = \sum_{k} U_{\al k} U_{\be k}^* \, e^{-iE_kL}.
\eeq
In this expression,
the mixing matrix $U_{\al k}$ and the eigenvalues $E_k$ of the unperturbed problem are necessary.
If conventional matter effects are negligible,
then the mixing matrix and the eigenvalues are simply given in terms of the of the mixing angles and mass-squared differences of the vacuum hamiltonian \eqref{H0}.

The second term in the series corresponds to the first modification due to the unconventional physics, 
which according to \eqref{CPT=NSI} could be either NSI or CPT violation.
For isotropic CPT violation,
this term has the explicit form \cite{DKM2009}
\beq\label{P1}
P_{\nu_\be\to\nu_\al}^{(1)} = 2L \, \text{Im} \B[S^{(0)*}_{\ab} \sum_{\ga\de}\M{\ab}{\ga\de}\,\a_{\ga\de}\B],
\eeq
where the sum over flavor indices shows that different components of the hamiltonian \eqref{H_CPT} can be studied with this method.
The complex weighting factors $\M{\ab}{\ga\de}$ are defined in terms of the unperturbed mixing parameters \cite{DKM2009}.
The probability \eqref{P1} is linear in the coefficient controlling CPT violation and arises due to the interference between the conventional hamiltonian and the perturbative correction \eqref{H_CPT}.
Making use of the relation \eqref{CPT=NSI},
the above expression for the probability can then be used to study the components of the NSI hamiltonian \eqref{H_NSI}.
The overall factor $L$ in the probability \eqref{P1} shows that the minute effects of the parameters $\a_{\ab}$ or $\e_{\ab}$ can be enhanced by a long propagation distance,
as expected from an interferometric measurement.

The second-order term in the perturbative expansion of the oscillation probability corresponds to two quadratic contributions \cite{DKM2009}.
The first of these effects is analog to the linear probability \eqref{P1} arising from the interference between CPT violation and the mass-driven hamiltonian.
The second quadratic effect is only due to CPT violation, 
leading to oscillations even if the conventional effects are negligible.
This is the leading-order term in the perturbative series for experimental configurations for which the zeroth-order amplitude is negligible.
This condition applies to experiments having short baseline compared to the mass-driven oscillation length,
either because the propagation distance is too short or because the neutrinos used have very high energy \cite{KM2004_SB}.

In addition to perturbative methods,
the exact treatment of the components of the coefficient $\a_{\ab}$ can also be implemented.
In fact, for experiments studying neutrinos over a large range of energies and baselines, 
such as atmospheric neutrinos, 
perturbative methods can become unapplicable. 
Exact diagonalization methods have been available in the literature since the realization of the importance of matter effects.
Many of these methods have been designed for long-baseline experiments,
in which case neutrinos travel for several hundreds of kilometers through rock \cite{Diag}.
For completeness,
a very compact method for the exact diagonalization of an arbitrary $3\times3$ hermitian hamiltonian as well as the general form of the mixing matrix are presented in Appendix \ref{Sec.ExDiag}.

\section{Bounds on NSI from CPT violation}
\label{Sec.bounds1}

Most of the searches for CPT violation in neutrinos have focused on coefficients that generate sidereal variations of the oscillation probability.
Only a few experiments have studied spectral distortions generated by isotropic coefficients for CPT violation.
The MiniBooNE collaboration obtained the $2\si$ bound $|\a_{e\mu}|<4.2 \times 10^{-20}\,\text{GeV}$
\cite{LV_MiniBooNE1,LV_MiniBooNE2} and the Double Chooz collaboration determined the limit 
$|\a_{e\tau}| < 7.8 \times 10^{-20}\,\text{GeV}$ at the 95\% C.L. \cite{LV_DC,LV_DC_KatoriSpitz}.
In both cases,
neutrinos propagate only a few hundred meters through the Earth's crust, 
where the matter density is about 1 gr/cm$^3$ \cite{PREM}.
Given the relatively short baseline of these two experiments,
their sensitivity is limited and the listed limits correspond to the bounds on NSI parameters
\beq\label{e_bounds1}
|\e_{e\mu}|  < 1.1 \times 10^{3}, \quad
|\e_{e\tau}| < 2.0 \times 10^{3}.
\eeq
These bounds are orders of magnitude weaker than the phenomenological limits already known in the literature \cite{Biggio:2009}.
Nevertheless,
it should be emphasized that the values \eqref{e_bounds1} are the first bounds on the magnitude of the $\e_{e\mu}$ and $\e_{e\tau}$ components directly obtained from an experimental analysis.

\begin{figure}[t]
\centering
\includegraphics[width=0.48\textwidth]{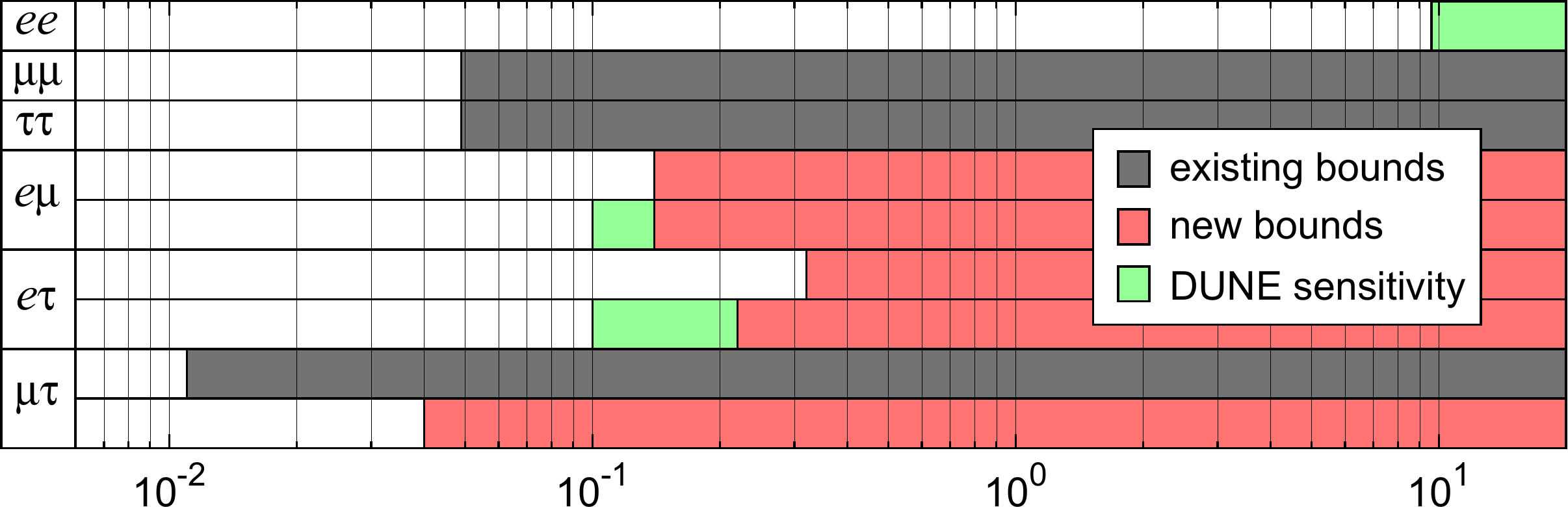}
\caption{Individual existing and new experimental bounds on the matter NSI parameters $\e_{\ab}$.
The two rows for the cases $\al\neq\be$ correspond to real and imaginary parts, respectively.
The estimated sensitivity of the future DUNE experiment is also shown.}

\label{Fig:e}
\end{figure}

In a recent study,
the Super-Kamiokande collaboration determined the most stringent upper bounds on real an imaginary parts of the off-diagonal components of the isotropic coefficient for CPT violation at the 95\% C.L. \cite{LV_SuperK}
\bea\label{aL_bounds_SK}
\R\a_{e\mu} &<& 1.8 \times 10^{-23}\,\text{GeV}, \nn\\
\I\a_{e\mu} &<& 1.8 \times 10^{-23}\,\text{GeV}, \nn\\
\R\a_{e\tau} &<& 4.1 \times 10^{-23}\,\text{GeV}, \nn\\
\I\a_{e\tau} &<& 2.8 \times 10^{-23}\,\text{GeV}, \nn\\
\R\a_{\mu\tau} &<& 6.5 \times 10^{-24}\,\text{GeV},\nn\\
\I\a_{\mu\tau} &<& 5.1 \times 10^{-24}\,\text{GeV}.
\eea
It is tempting to simply apply the relation \eqref{CPT=NSI} to translate the bounds on $\a_{\ab}$ into $\e_{\ab}$;
however,
given the multiple directions and creation points of atmospheric neutrinos used for the analysis, 
the matter potential experienced by these neutrinos is far from unique.
A modest estimate can be obtained by using an appropriate average density for all the neutrino events.
The data analysis found that up-going events are the most sensitive sample to the unconventional effects studied,
which is expected due to the large travel path of these neutrinos.
Since 90\% of up-going neutrinos propagate through matter with an average density of 3.4 gr/cm$^3$ or more \cite{PREM},
we can take this value as a moderate approximation for the average density of matter used to determine the bounds \eqref{aL_bounds_SK}.
We remark that this is a conservative estimate because a significant fraction of the events used in the analysis traverse denser layers of the Earth that would lead to even more stringent bounds on the NSI parameters.
The relation \eqref{CPT=NSI} then gives the following upper bounds on the real and imaginary parts of the NSI parameters
\begin{align}\label{e_bounds_SK_new}
\R\,\e_{e\mu} &< 1.4 \times 10^{-1}, 
&\I\,\e_{e\mu} &< 1.4 \times 10^{-1}, \nn\\
\R\,\e_{e\tau} &< 3.2 \times 10^{-1}, 
&\I\,\e_{e\tau} &< 2.2 \times 10^{-1}, \nn\\
\R\,\e_{\mu\tau} &< 5.1 \times 10^{-2},
&\I\,\e_{\mu\tau} &< 4.0 \times 10^{-2}.
\end{align}
It should be noticed that the values \eqref{e_bounds_SK_new} include the first bounds on the imaginary parts of these off-diagonal parameters.
Until now,
only the magnitude of these parameters has been considered.
Additionally,
these bounds are the first on the $\e_{e\mu}$ and $\e_{e\tau}$ components directly obtained from an experimental analysis.
Even considering indirect phenomenological bounds on these components,
the values \eqref{e_bounds_SK_new} exhibit a slight refinement on $|\e_{e\mu}|$,
and improvements by more than a factor five on $|\e_{\mu\tau}|$ 
and almost one order of magnitude on $|\e_{e\tau}|$.
Fig. \ref{Fig:e} illustrates existing bounds as well as the improvements obtained in this work by considering that only one of the parameters at a time is nonzero.

\section{Bounds on CPT violation from NSI}
\label{Sec.bounds2}

Systematic experimental searches of matter NSI have been only performed to study the $\mu\tau$ sector.
Using accelerator neutrinos, 
the MINOS collaboration \cite{NSI_MINOS}
implemented a two-flavor parametrization of NSI and using both neutrino and antineutrino data
determined a bound on the real part of $\e_{\mu\tau}$ at the 90\% C.L. given by 
\beq\label{e_bound_MINOS}
-0.20 < \R\,\e_{\mu\tau}  < 0.07.
\eeq
Reaching a maximum depth of about 10 km,
more than 81\% of the 735-km path of the neutrino beam of the MINOS experiment goes through rock of 
density 2.6 gr/cm$^3$ \cite{PREM}.
Thus, 
the two-sided bound \eqref{e_bound_MINOS} translates into a limit on the coefficient for CPT violation
\beq\label{aL_bound_MINOS_new}
-1.95\times 10^{-23}\,\text{GeV} 
< \R\a_{\mu\tau}  
< 6.83\times 10^{-24}\,\text{GeV}.
\eeq
The remaining 19\% of the beam propagates through less dense matter;
however,
we have only kept the most conservative value.
Notice that the upper bound is slightly weaker than the limit shown in \eqref{aL_bounds_SK};
nevertheless,
the result from Super-Kamiokande is one sided, 
whereas the result \eqref{aL_bound_MINOS_new} includes the first lower bound on $\R\a_{\mu\tau}$.

\begin{figure}[t]
\centering
\includegraphics[width=0.48\textwidth]{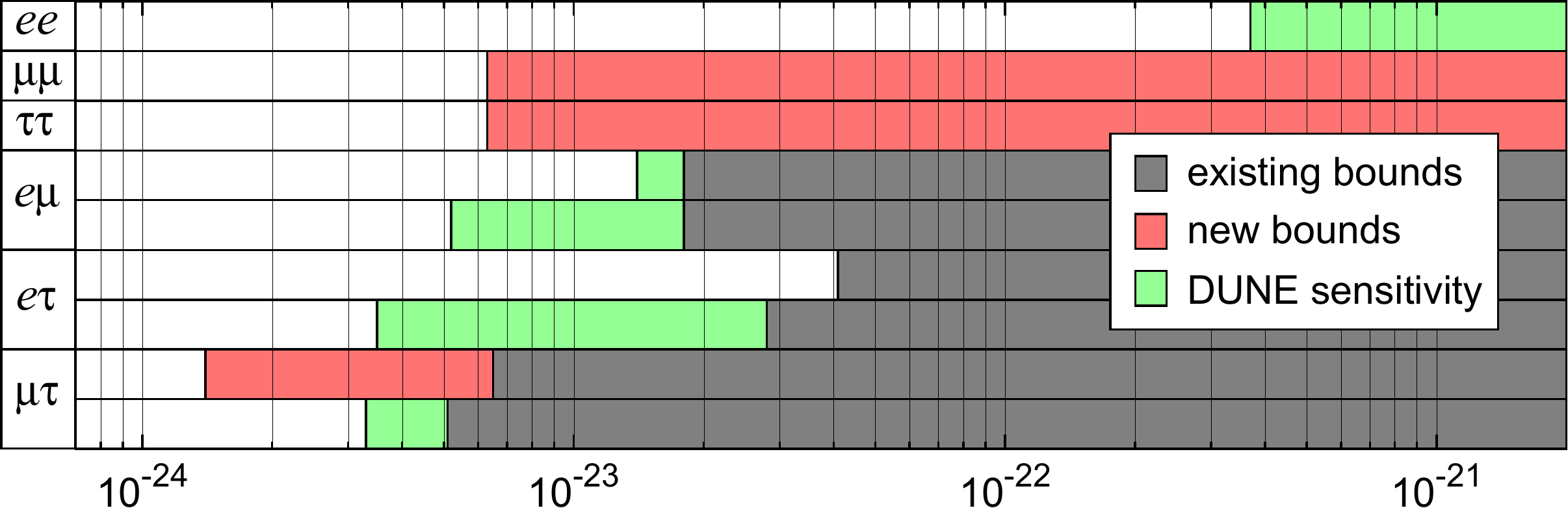}
\caption{Individual existing and new experimental bounds on the coefficients for CPT violation $\a_{\ab}$ in GeV.
The two rows for the cases $\al\neq\be$ correspond to real and imaginary parts, respectively.
The estimated sensitivity of the future DUNE experiment is also shown.} 
\label{Fig:aL}
\end{figure}

Similarly,
using atmospheric neutrinos and a two-flavor approximation for the characterization of NSI, 
the Super-Kamiokande collaboration determined the following bounds in the $\mu\tau$ sector at the 90\% C.L. \cite{NSI_SuperK} 
\bea\label{e_bounds_SK}
|\R\,\e_{\mu\tau}|  &<& 1.1\times10^{-2},
\nn\\
|\e_{\mu\mu}-\e_{\tau\tau}| &<& 4.9\times10^{-2}.
\eea
Following the approach presented in Sec. \ref{Sec.bounds1},
atmospheric neutrinos propagate through matter of different density;
therefore,
an appropriate application of the correspondence in Eq. \eqref{CPT=NSI} requires the use of a reasonable value for an average density.
In the same conservative approximation used in Sec. \ref{Sec.bounds1},
the two-sided bounds \eqref{e_bounds_SK} translate into the following limits on the corresponding coefficients for CPT violation
\bea\label{aL_bounds_SK_new}
|\R\a_{\mu\tau}|  &<& 1.4\times 10^{-24}\,\text{GeV},
\nn\\
|\a_{\mu\mu}-\a_{\tau\tau}| &<& 6.3\times10^{-24}\,\text{GeV}.
\eea
In addition of being two-sided bounds,
the above result provides a slight improvement on $\R\a_{\mu\tau}$
and the first limit involving diagonal components.
Following the standard method for estimating attainable sensitivities, 
we consider one coefficient at a time to write the following bounds on the relevant diagonal components
$|\a_{\mu\mu}| < 6.3\times10^{-24}\,\text{GeV}$ and
$|\a_{\tau\tau}| < 6.3\times10^{-24}\,\text{GeV}$.
Fig. \ref{Fig:aL} illustrates existing bounds as well as the improvements obtained in this work by considering that only one of the parameters at a time is nonzero.

\section{Future prospects}
\label{Sec.future}

Since the neutrino propagation distance enhances the minute effects of $\e_{\al}$ and $\a_{\al}$,
several current and future experiments,
such as 
the Deep Underground Neutrino Experiment (DUNE) \cite{DUNE},
NuMI Off-Axis $\nu_e$ Appearance experiment (NO$\nu$A) \cite{NOvA},
and 
the Precision IceCube Next Generation Upgrade (PINGU) \cite{PINGU}
offer exciting prospects for the study of matter NSI and CPT violation.

As an illustration, 
we can determine an estimate on the sensitivity of DUNE.
A comprehensive sensitivity assessment can be done using GLoBES \cite{GLoBES};
however, 
here we simply consider a crude estimate in the form of a 10\% deviation from the conventional oscillation probability.
DUNE is a multipurpose experiment designed for a rich physics program including the study of long-baseline neutrino oscillations, 
supernova neutrinos, and
atmospheric neutrinos.
As a long-baseline experiment,
improvements on the current bounds are expected due to the 1300 km. that separate the source and the far detector.
Sensitivities to the matter-NSI parameters $\e_{\ab}$ are presented in Fig. \ref{Fig:e},
while the corresponding sensitivities to the coefficients for CPT violation $\a_{\ab}$ are presented in Fig. \ref{Fig:aL}.
In both cases,
the potential bounds are illustrated by considering that only one of the parameters at a time is nonzero.

For matter NSI,
we find that the bounds from CPT violation obtained in this work filled most of the previously unexplored parameter space with great sensitivity.
Nonetheless,
DUNE could improve the bounds on the electron sector.
For CPT violation,
bounds on all the coefficients in the electron sector could be improved by DUNE as well as the imaginary part of $\a_{\mu\ta}$.

\section{Summary}
\label{Sec.summary}

Neutrino oscillations offer a remarkable sensitivity to search for new physics.
Two particular descriptions of unconventional physics have been explored and shown to produce similar modifications.
Despite the completely different theoretical motivations and underlying physics,
nonstandard matter interactions and CPT violation, 
parametrized by the matrices $\e_{\ab}$ and $\a_{\ab}$, 
respectively, 
lead to the same observable effects. 
Since the data analysis only makes use of the oscillation probabilities,
from an experimental point of view the results from one framework are equivalent to the other.
This equivalence is explicitly presented in Eq. \eqref{CPT=NSI},
which has been used to relate the current bounds the parameters on both formalisms.

Under mild assumptions,
new limits have been determined in these sets of parameters derived from a systematic experimental analysis.
Despite the stringent phenomenological bounds on 
$|\e_{e\mu}|$, 
$|\e_{e\tau}|$, 
and $|\e_{\mu\tau}|$,
in Eq. \eqref{e_bounds_SK_new} these limits have been improved and bounds on the imaginary parts of these parameters have been constrained for the first time.
Similarly,
an increased sensitivity and a lower bound on the coefficient $\a_{\mu\tau}$ is found in Eq. \eqref{aL_bounds_SK_new} 
as well as the first bound on the combination of diagonal components $|\a_{\mu\mu}-\a_{\tau\tau}|$.

These results demonstrate the functionality of the correspondence between formalisms presented in this work.
Future bounds on coefficients for CPT violation in neutrino oscillations can be used to report bounds on matter NSI parameters, and vice versa.
In the case of a positive signal,
a distinction between matter NSI and CPT violation can be made because the former can only occur for neutrinos that propagate through a dense medium,
whereas the observable effects of the latter can occur in vacuum.

The vast experimental program in neutrino oscillations offers exciting opportunities for the study of neutrinos as well as the search of unconventional physics.
In the present work we have established a direct correspondence between two communities in the hunt for new physics using neutrinos.

\section*{Acknowledgments}
The author thanks C. Arg\"uelles and T. Katori for valuable comments.
This work was supported in part by the German Research Foundation (DFG) under Grant No. KL 1103/2-1.

\appendix
\section{Exact diagonalization}
\label{Sec.ExDiag}
We begin with the $3\times3$ hermitian hamiltonian in flavor basis, whose entries will be denoted by $H_{\ab}$.
Denoting the three eigenenergies by $E_j$ and the identity matrix by $\pmb{1}$, the eigenvalue equation $\text{det}(\pmb{H}-E_j\pmb{1})=0$ leads to the cubic equation that can be nicely written in terms of the hamiltonian invariants 
\beq\label{cubic}
E_j^3+M_{T} E_j^2+M_{TT} E_j + M_D = 0,
\eeq
where
\bea\label{Ms}
M_{T\phantom{T}} &=& - \Tr \,\pmb{H},
\nn\\
M_{TT} &=& \ha(\Tr\,\pmb{H})^2- \ha\Tr(\pmb{H}^2),
\nn\\
M_{D\phantom{T}} &=& - \text{det}\,\pmb{H}.
\eea

\ni
The matrix invariants are functions of the entries of the hamiltonian, which can be used to implement Cardano's method for solving the cubic equation \eqref{cubic}.
By defining the angles
\beq
\th_j = \!\frac{1}{3}\left[\arccos{\!\left(\!\frac{2M_T^3-9M_T M_{TT}+27M_D}{2(M_T^2-3M_{TT})^{3/2}}\right)}\!+t_j\right],
\eeq
with $t_1=0,\,t_2=-t_3=2\pi$,
the three solutions of the cubic equation \eqref{cubic} can be written as 
\beq\label{E_j}
E_j = -\frac{1}{3}\B[2(M_T^2-3M_{TT})^{1/2}\,\cos\th_j + M_T\B].
\eeq

Once the eigenenergies have been found, we can use the eigenvalue equation to construct the eigenstates, which will constitute the columns of the mixing matrix.
The nine entries of the mixing matrix can be explicitly written as
\beq\label{U_alj}\!
U_{ej} = \frac{|\al_j\be_j|}{N_j}, \;\;
U_{\mu j} = \frac{\al_j\ga_j}{N_j}\frac{|\be_j|}{|\al_j|}, \;\;
U_{\ta j} = \frac{\be_j\ga_j^*}{N_j}\frac{|\al_j|}{|\be_j|}, 
\eeq

\ni
where we have defined the functions
\bea
\al_j&=&H_{\tau e}\BB(H_{\mu\mu}-E_j\BB)-H_{\ta\mu}H_{\mu e} ,
\nn\\
\be_j&=&H_{\mu e}\BB(H_{\tau\tau}-E_j\BB)-H_{\mu\tau}H_{\tau e} ,
\nn\\
\ga_j&=&H_{\mu\tau}\BB(H_{ee}-E_j\BB)-H_{\mu e}H_{e\ta} ,
\eea

\ni
and the normalization factors
\beq
N_j = \BB(|\al_j\be_j|^2+|\al_j\ga_j|^2+|\be_j\ga_j|^2\BB)^{1/2}.
\eeq

\ni
Now that we have explicit forms for the eigenenergies \eqref{E_j} and the elements of the mixing matrix \eqref{U_alj}, we can use the general form of the oscillation probability 
\bea
P_{\nu_\al\to\nu_\be} &=& \de_{\ab}-
4\sum_{j>k} \R \BB(U_{\al j}^*U_{\be j}U_{\al k}U_{\be k}^*\BB) \, \sin^2\!\BB(\ha\De_{jk}t\BB)
\nn\\
&& + 2\sum_{j>k} \I \BB(U_{\al j}^*U_{\be j}U_{\al k}U_{\be k}^*\BB) \, \sin(\De_{jk}t)
\eea
to write the probabilities of interest,
where the eigenvalue differences are $\De_{jk}=E_j-E_k$.


%
\end{document}